\documentclass[twocolumn,showpacs,preprintnumbers,amsmath,amssymb]{revtex4}
\usepackage{psfig}
\usepackage{color}
\usepackage{dcolumn}
\usepackage{amsmath}
\begin{document}

\title{Electronic structure and non-magnetic character of $\delta$-Pu-Am alloys}

\author{Alexander Shick$^1$, Ladislav Havela$^2$, Jind\u{r}ich Koloren\u{c}$^1$, V\'aclav Drchal$^1$,
Thomas Gouder$^3$,  Peter M. Oppeneer$^4$}
\address{$^1$\ Institute of Physics, ASCR, Na Slovance 2, CZ-18221 Prague
8, Czech Republic \\
$^2$Charles University, Faculty of Mathematics and Physics,
Department of Electronic Structures, Ke Karlovu 5, CZ-12116 Prague
2, Czech Republic\\
$^3$European Commission, Joint Research Centre, Institute for
Transuranium Elements, Postfach 2340, 76 125 Karlsruhe, Germany \\
$^4$Department of Physics, Uppsala University, Box 530, S-751 21
Uppsala, Sweden}

\date{\today}

\begin{abstract}
The  {\em around-mean-field} LSDA+U correlated band theory is
applied to investigate the electronic and magnetic structure of
$fcc$-Pu-Am alloys. Despite a lattice expansion caused by the Am
atoms, neither tendency to 5$f$ localization nor formation of local
magnetic moments on Pu atoms in Pu-Am alloys are found. The
$5f$-manifolds in the alloys are calculated being very similar to a
simple weighted superposition of elemental Pu and Am $5f$-states.
\end{abstract}
\maketitle

Electron-electron correlation effects in a proximity to the
localization threshold of the $5f$ series, which is crossed between
Pu and Am, recently attracted significant attention in condensed
matter physics \cite{Kotliar,grivenau,heathman}. While strongly
correlated systems usually have local magnetic moments, the whole
set of experimental data clearly demonstrates their absence in Pu
\cite{lashley}. Recent LSDA+U theoretical calculation \cite{shick05,Shorikov}
have explained the non-magnetic character of $\delta$-Pu, leading to
the 5$f^{6}$-like state, hybridized with conduction-electron states.
An important issue remains open whether a further $5f$ localization
or often speculated onset of magnetism can take place when expanding
the lattice by Am doping.

Alloying Pu with Am provides an important opportunity to stabilize
the high-temperature $\delta$-phase down to $T= 0$. Since Am atoms
are larger than Pu, they create a ``negative" pressure increasing
the Pu interatomic spacing. Naively, knowing the sensitivity of the
Pu phases to the lattice spacing, one can expect the formation of
local magnetic moments at Pu due to an increase of $5f$-states
localization.
In contrast, recent
susceptibility measurements of the Pu-Am systems do not show any
systematic variations and susceptibility remains very weakly
temperature dependent, not exceeding 1$\cdot$10$^{-8}$ m$^{3}$/mol
\cite{baclet}. Also results of specific-heat studies \cite{javorsky}
indicate no sign of magnetism in Pu-Am alloys.

The Pu-Am alloys were studied theoretically in Ref.
\cite{landa} using the local spin density (LSDA) and  generalized
gradient (GGA) approximations to the Density Functional theory
(DFT), and no electron-electron correlations beyond those
incorporated in LSDA/GGA were included. Large local magnetic moments
on Pu and Am atoms in Pu-Am alloys (disordered or
antiferromagnetically coupled) were calculated in Ref.
\cite{landa}, in contradiction to experimental findings
\cite{baclet,javorsky}. The situation is similar to the DFT
calculations (both within LSDA and GGA approximations) for elemental
$\delta$-Pu \cite{soderlindPu,niklasson} and $fcc$-Am
\cite{soderlindAm1,soderlindAm2}, where magnetism tends to appear as
an undesirable artifact.

In this Communication we go beyond the conventional LSDA/GGA
approximations and apply the LSDA+U correlated energy-band approach
to investigate the electronic structure of the Pu-Am alloys. The
LSDA+U method \cite{AZA91,LAZ95} starts from the LSDA total energy,
which is supplemented by an additional intra-atomic Coulomb $U$
correlation term and an intraatomic exchange interaction $J$ term of
multiband Hubbard-type minus a so-called double counting term, to
subtract the electron-electron interaction already included in the
LSDA .
Our calculations are
performed using the relativistic version of the {\em
``around-mean-field"} LSDA + U method (AMF-LSDA+U) \cite{shick05},
as implemented in the full-potential linearized augmented-plane-wave
(FP-LAPW) basis \cite{shick99} in which the spin-orbit coupling
(SOC) is taken into account \cite{shick01}.

Before considering in detail the effect of Coulomb correlations  for
the Pu-Am alloys, we need to make sure that the AMF-LSDA+U is
capable to deal with the end-points, namely $\delta$-Pu and
$fcc$-Am. Applied to ${\delta}$-Pu \cite{shick05}, this approach
accounted for all basic characteristics: equilibrium volume, bulk
modulus, $5f$-manifold binding energy, including the disinclination to
magnetic ordering, appearing due to a combined effect of SOC and
Coulomb $U$, which lead to a nearly closed $5f_{5/2}$ sub-shell,
hybridized with conduction electron states.
We note that the non-magnetic $\delta$-Pu can be obtained in fully-localized-limit
(FLL)-LSDA+U calculations \cite{Shorikov} when no spin-polarization is allowed in the LSDA part
and unreasonably small value of the exchange $J \; < \; 0.48$ eV is used.

Here we focus on the issue how can the same method describe similar
(in terms of the $5f$ occupancy), but presumably much more localized
$5f$-states in Am metal. We restricted ourselves to the $fcc$ phase,
which covers a broad concentration range from $\delta$-Pu nearly to
pure Am \cite{ellinger}. Also, it is the Am phase stable at elevated
temperatures and at high pressure \cite{heathman00}. The Am atom has
six $f$ electrons with full $j=5/2$ shell and zero magnetic moment.
When the $5f$ states are treated as non-spin-polarized valence
states in GGA, the calculated equilibrium volume $V_{eq} \approx
120$ (a.u.)$^3$ \cite{pen02} is much lower (almost by 40 \%) than
the experimental value of $198$ (a.u.)$^3$ \cite{lindbaum}. When
spin polarization is included in GGA calculations
\cite{soderlindAm1}, augmented by well known orbital polarization
correction \cite{brooks} (GGA+OP), the $V_{eq} = 170$ (a.u.)$^3$
improves significantly. However, large spin $M_s$ = 6 $\mu_B$ and
orbital $M_l$ = $-$1 $\mu_B$ moments are induced at the Am atom.
This strong magnetic state, either ordered or disordered, was never
seen experimentally.

Another line to improve $V_{eq}$ assumes the $5f$-manifold to be a
part of an atomic-core (the ``open-core" approximation), and
neglects an overlap between the $5f$-states and the rest of valence
electrons. It increases the volume to 180 \cite{soderlindAm1} or 202
\cite{pen02} (a.u.)$^3$ in agreement with the experiment. While the
``open-core" approximation is based on a very strong constraint and
corresponds to a non-physical limit of infinitely large Coulomb
repulsion in the f-shell, it illustrates the impact of the
$5f$-localization for Am. Alternatively,  self-interaction
correction (SIC) to LSDA can be employed to localize the $5f^6$
manifold. The LSDA-SIC yields the $V_{eq}$ of $213$ (a.u.)$^3$
\cite{petit00} which is somewhat larger than experimental value.
Similarly to the ``open-core", the LSDA-SIC gives a localized Am
$5f^6$ shell. While being {\em ab-initio} and working well at Am
end, the LSDA-SIC strongly overestimates the localization in
$\delta$-Pu yielding $V_{eq}=218$ (a.u.)$^3$ \cite{petit00}
exceeding the experimental value by about 30 \%. This illustrates
the difficulty of LSDA-SIC method to deal with electron-electron
correlations in $\delta$-Pu, which limits further the applicability
of LSDA-SIC to Pu-Am alloys.

We performed AMF-LSDA+U calculations for Am, starting from the
strongly spin-polarized LSDA charge and spin densities and on-site
spin and orbital occupations, and without any constraints. We varied
the Coulomb $U$ from 3 eV to 4 eV, and chose the intra-atomic
exchange parameter $J$=0.75 eV, i.e. in the range of commonly
accepted values for Am. The calculations converged to practically
zero magnetic moment with residual $M_s$ and $|M_l|$ less than 0.01
$\mu_B$, and $5f$-manifold occupation  $n_{5f} = 5.9$. Also, $V_{eq}
= 186$ (a.u.)$^3$ and bulk modulus $B = 55.1$ GPa are calculated for
$U$ = 4 eV, in a reasonable agreement (considering the different
structure type) with experiment (see Table I.).

\begin{table}[floatfix]
\vspace*{-0.25cm}
\caption{Ground state properties of $fcc$-Am. Given are the $5f$
occupation number $n_{5f}$, the spin moment $M_s$, the orbital
moment $M_l$, and the total magnetic moment $M_{j}$ (spin+orbital
), and equilibrium volume ($V_{eq}$) and bulk
modulus ($B$). The experimental value of $B$, taken from
Ref. \cite{lindbaum}, belongs to the $dhcp$ structure.}
\begin{tabular}{ccccccc}
\hline
Method & $n_{5f}$  & $M_s$, $\mu_B$ & $M_l$, $\mu_B$ & $M_{j}$, $\mu_B$& $V_{eq}$,  a.u.$^3$& $B$, GPa \\
\hline
GGA+OP \cite{soderlindAm1}& N/A & 6      &  -1   &  5   & 170 & 43.0 \\
open-core \cite{pen02}& 6 & 0      &   0   &  0   & 202 & 46.8 \\
open-core \cite{soderlindAm1} & 6 & 0      &  0   &  0   & 180 & 46.0 \\
LSDA-SIC \cite{petit00}  & 6 & 0      &  0   &  0   & 213 & 39.5 \\
\hline
LSDA+U=3 eV & 5.92  & 0 & 0 & 0  & 182 & 67.3  \\
LSDA+U=4 eV & 5.90  & 0 & 0 & 0  & 186 & 55.1  \\
Exp. \cite{lindbaum} & 6     & N/A & N/A & 0 & 198 & 29.9 \\
\hline
\end{tabular}
\label{table1}
\end{table}

\begin{figure}[floatfix]
\vspace*{-0.1cm}
\centerline{\psfig{file=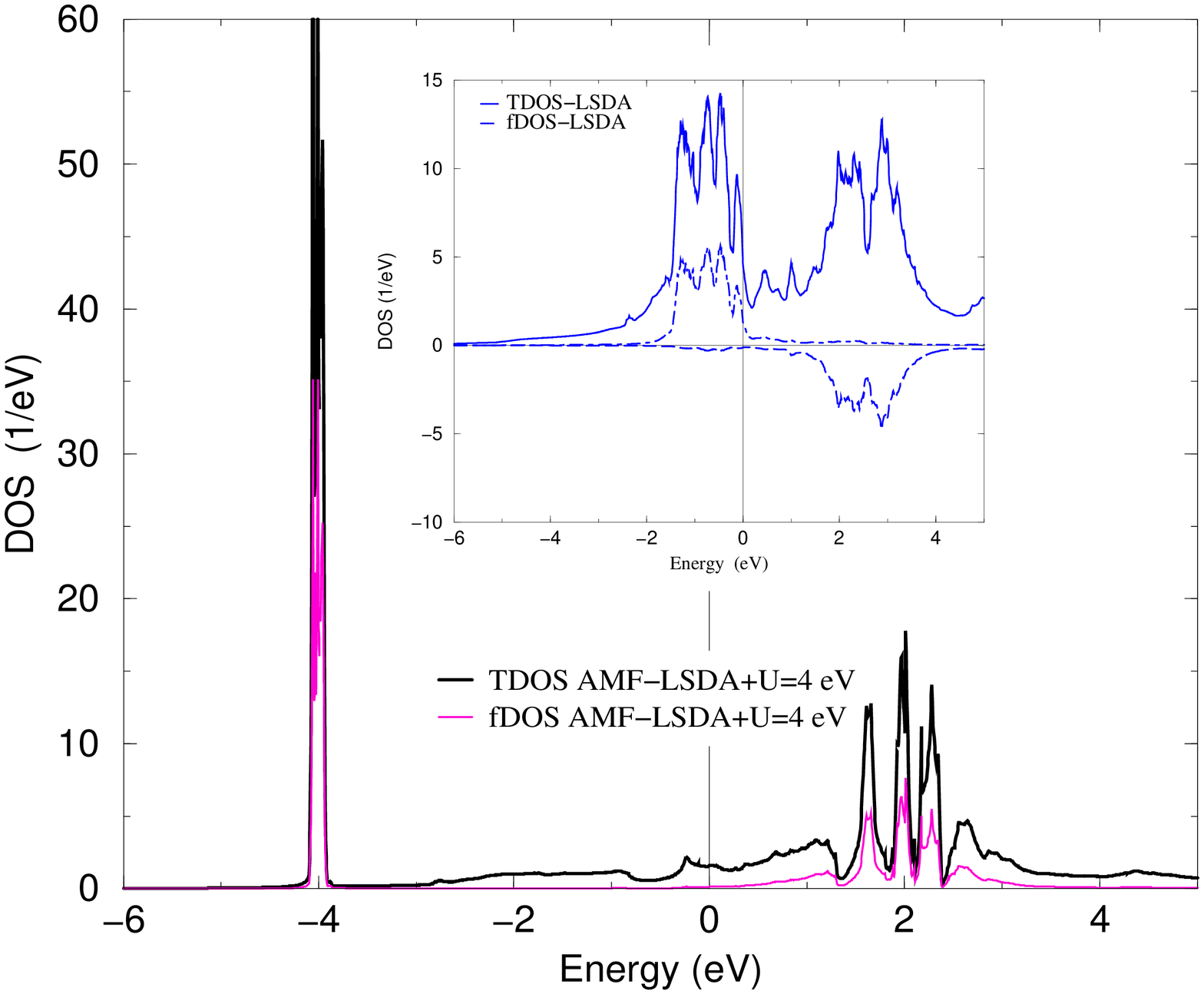,width=7cm,height=6.0cm}}
\vspace*{-0.3cm}
\centerline{\psfig{file=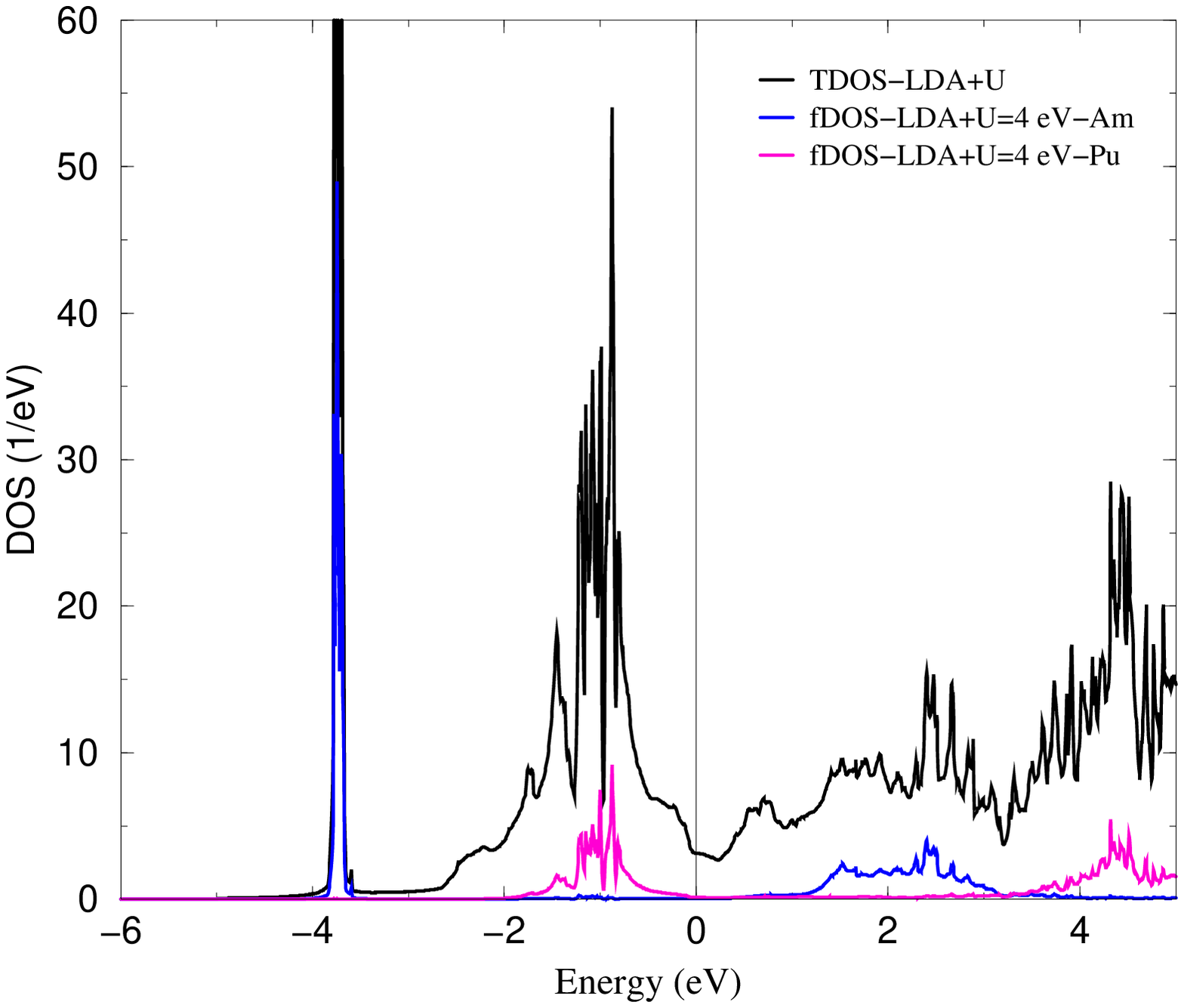,width=7cm,height=6.5cm}}
\vspace*{-0.4cm} \caption{Color-on-line. (top) The DOS (TDOS-total
and fDOS-projected) for \textit{fcc}-Am from LSDA (inset), and
AMF-LSDA+U= 4 eV calculations.
(bottom) The TDOS and fDOS
for Pu$_3$Am alloy}
\end{figure}

In Fig. 1 (top) we show the total (TDOS) and $f$-states projected
densities of states (fDOS) resulting from the $fcc$-Am LSDA and
AMF-LSDA+U=4 eV calculations.  The strong spin-polarization found in
LSDA disappears in AMF-LSDA+U, resulting in a truly non-magnetic
($S_z=L_z=J_z=0$) $5f^6$ ground state.  The $5f$ manifold binding
energy is calculated as $\approx$ 3.5 eV (for $U=$ 3 eV), 4 eV ($U=$
4 eV). The LSDA+U method is not based on any kind of atomic coupling
scheme ($LS$ or $jj$), it rather determines the set of
single-particle orbitals that minimize variationally the total
energy. The AMF-LSDA+U result can be interpreted as yielding the
$J=0$ singlet ground-state configuration that corresponds to the
$jj$-coupled Slater determinant formed of fully and equally
populated six $j=5/2$ orbitals.

Although the DOS cannot be directly compared with $5f$ features in
valence-band photoelectron spectra (PES), as - for localized $5f$
states - those reflect not the $5f^{n}$ ground state, but the whole
$5f^{n-1}$ multiplet, the energy range is undoubtedly correct
\cite{naegele,gouder05}. Comparing to results of Pu calculations
(see Fig. 3 of Ref. \cite{shick05}), the Am $5f$ states are not only
at higher binding energy, but also conspicuously narrower, with the
width not exceeding 0.1 eV.
The results obtained for Am at ambient
pressure can be definitely characterized as $5f$-localized.
Shift of
the $5f$ states in AMF-LSDA+U down from the vicinity of the Fermi
energy ($E_F$) decreases their contribution to the chemical bonding
and increases  $V_{eq}$.

The situation of the $5f$ states changes dramatically when the volume
is reduced from 198 (a.u.)$^{3}$ to 133 (a.u.)$^{3}$, which corresponds
to the volume of the high pressure phase AmIV. This phase is
generally assumed, as the orthorhombic structure reminiscent of
$\alpha$-U, to have itinerant $5f$ states \cite{heathman00}. Indeed, our
calculations exhibit the $5f_{5/2}$ band broadened nearly up to 1
eV. The width of the $5f_{5/2}$ band becomes thus comparable to
$\delta$-Pu, although there remains a large difference in binding
energies: in $\delta$-Pu, the $5f_{5/2}$ band is located around 1 eV
below $E_F$, while in Am it is centered at 3.5 eV  even for the
high-pressure phase, just below the lower edge of the valence band.

Next, we turn to the salient aspect of our investigation, the
AMF-LSDA+U  calculations of Pu$_x$Am$_{1-x}$ (x = 0.25, 0.5, 0.75)
alloys. In all calculations, the Coulomb $U$= 4 eV, and exchange
$J$=0.75 eV for Am atom, and $U$=4 eV, $J$=0.70 eV for Pu atom were
used. The $fcc$-based super-cells were used, $L1_2$ for Pu$_3$Am
 (Pu$_{0.75}$Am$_{0.25}$) and  PuAm$_{3}$  (Pu$_{0.25}$Am$_{0.75}$),
and $L1_0$  for PuAm  (Pu$_{0.5}$Am$_{0.5}$). The lattice
parameters, 8.905 (a.u.)$^3$ for Pu$_{0.75}$Am$_{0.25}$, 9.018
(a.u.)$^3$ for Pu$_{0.5}$Am$_{0.5}$, 9.132 (a.u.)$^3$ for
Pu$_{0.25}$Am$_{0.75}$, were taken from Ref. \cite{ellinger}.

The LSDA calculations produce magnetic solutions for all considered
Pu-Am alloys, similar to those reported in Ref. \cite{landa}.
Starting from these spin-polarized LSDA charge and spin densities
and on-site spin and orbital occupations, we apply AMF-LSDA+U. The
magnetism collapses in all cases, and Pu-Am alloys become
non-magnetic, similarly to $fcc$-Am and $\delta$-Pu. The calculated
DOS (total per unit cell and $5f$-projected (fDOS)) are shown in
Fig.1 (bottom) for Pu$_{0.75}$Am$_{0.25}$ (Pu$_3$Am) alloy. The Am
$5f$-manifold is located at $\sim$4 eV below $E_F$, similarly to
AMF-LSDA+U DOS for $fcc$-Am (see Fig. 1(top)). The Pu-$5f$ states
are located in a region from -2 eV to 0.5 eV and peaked at $\sim$1
eV below $E_F$, similarly to the DOS of $\delta$-Pu shown in Fig. 3
of Ref. \cite{shick05} (one also should keep in mind that there are
small DOS changes due to the change in the lattice parameter).
Importantly, the alloy DOS is almost exactly a weighted
superposition of pure $fcc$-Am and $\delta$-Pu DOS showing that
there is a very little interaction between Am- and Pu-$5f$
manifolds.  The calculated DOS for Pu$_{0.5}$Am$_{0.5}$ and
Pu$_{0.25}$Am$_{0.75}$ look qualitatively very similar to
Pu$_{0.75}$Am$_{0.25}$ DOS shown in Fig. 1 (bottom), with well
separated Am and Pu $5f$-manifolds, whose relative heights are
proportional to the Am and Pu atom concentration, again supporting
the smallness of the interaction between the Am- and Pu-$5f$ states.
The difference between equilibrium volumes of $\delta$-Pu and
$fcc$-Am is relatively small ($\approx$ 15 \%) and no significant
deviation from Vegard's law is found in experiment \cite{ellinger}.
Considering the lattice compression with a decrease of Am content,
we did not find  any sizeable effect on Am-$5f$ states. The lattice
compression is simply too small compared to the changes at hight
pressure, and Am remains still practically in the AmI phase, i.e.
below the first phase transition at 6 GPa. Similarly, there is no
significant change in Pu-$5f$ states found with a decrease of Am
content, just a small ($\approx$ 0.2 eV) decrease in the binding
energy. It is also noteworthy that we did not find any sizable
change in the Pu-$5f$ and Am-$5f$ manifold occupations ($n_{5f}$)
\cite{occupation}.

\begin{figure}[floatfix]
\vspace*{-0.5cm}
\psfig{file=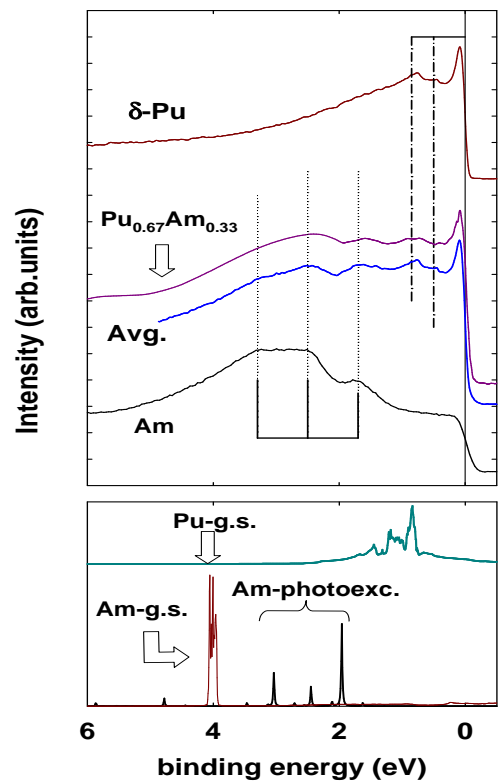,width=9.5cm,height=11cm}
\vspace*{-0.4cm}
\caption{Color-on-line. Upper panel:valence-band photoelectron
spectra (40.81 eV photon energy) for Am and $\delta$-Pu.
The lower panel displays the calculated ground-state DOS for Am and
$\delta$-Pu and the position of excitations obtained from the atomic
calculations of Am (black). }
\end{figure}

In Fig.2 we show the valence-band PES of $\delta$-Pu, Am, and
Pu$_{0.67}$Am$_{0.33}$, obtained with the photoexcitation energy
40.81 eV, emphasizing the emission from the $5f$ states. Although in
the case of band systems PES provides practically a snapshot of the
density of one-electron states, final-state effects are important
for localized as well as narrow-band systems. In analogy to
rare-earths, the valence-band spectrum of Am should exhibit the
multiplet of excited states. To address these effects, the
one-particle Green's function and corresponding spectral density
were calculated for Am atom. In these calculations we used exact
diagonalization of the many-body hamiltonian for $f^n$-shell closely
following Ref. \cite{LK99}, adding the SOC ($\xi$=0.346 eV), and
using the values $U$= 4 eV and $J$= 0.75 eV, as in the AMF-LSDA+U
calculations. Also, $k_BT$ = 0.1 eV and $\langle \emph{N}_{f}
\rangle=$ 6 were used to determine the chemical potential ($E_F$).
The calculated spectral density shown in Fig. 2 (Am - photoexcitations)
and agrees very well with similar recent calculations of Svane \cite{svane}.
It is
in a conspicuously good agreement with experimental PE spectra. Such
comparison cannot be done in the case of Pu, as the use of the
atomic limit for the Pu $5f$ states is not justified. The Am-$f^6$
many-body ground state is found to have $J=0$ character with both
$S=L=2.471$, while neither $S$ nor $L$ is a good quantum number. It
differs from the the result of Ref. \cite{savrasov05} that yields
LS-coupled $^7F_0$ ground state.

Although the experimental PES cannot be accounted for by the LSDA+U
theory, one can see that the calculated spectrum of the Pu-Am alloy
corresponds to the weighted average of spectra of pure Pu and Am,
similarly to the PE spectra.  As shown recently \cite{gouder},  the
Pu-$4f$ spectra of Pu-Am alloys remain practically identical
irrespective of Am concentration at least up to $\approx$ 35 $\%$
Am. Also, very recent results of specific-heat studies
\cite{javorsky} indicate no enhancement of the $\gamma$ coefficient
of the electronic specific heat.  We can conclude that the lattice
expansion caused by Am is not driving the Pu $5f$ states to further
localization and no magnetic moments on Pu in $\delta$-Pu-Am alloys
are induced thereby. Both from the experiment and theory, we deduce
that the character of the Pu-$5f$ states does not change with the Am
doping. The experimental PES features close to $E_F$ can not be
explained in terms of simple band states, similarly to the case of
$\delta$-Pu for which the dynamical effects  become very important
\cite{Kotliar}.

To summarize, we demonstrated that the {\em around-mean-field}
version of the LSDA+U method gives a unified picture of the
electronic structure of Am, $\delta$-Pu and $\delta$-Pu-Am alloys.
For $fcc$-Am  our calculations yield a non-magnetic ground state
which is almost exactly a localized $f^6$ $J = 0$ singlet state,
with equilibrium volume and bulk modulus in a good agreement with
experimental data. Based on the calculations, we expect that this
non-magnetic state will be preserved even for the
pressure-delocalized state of Am, in accord with electrical
resistivity \cite{grivenau} lacking any clear anomaly of magnetic
origin.

Neither magnetism nor further localization of the Pu-$5f$ states is
found in \textit{fcc} Pu-Am alloys on the LSDA+U level, i.e. without
introducing any additional effects like magnetic moment disorder or magnetic
fluctuations, which are disproved recently by $\mu$SR experiments  \cite{muons}.
We are aware that our static mean-field approach does not account for a possible
Kondo screening of the magnetic moment fluctuations,
which are important in other f-electron systems (e.g. $\alpha$-Ce).

The research was carried out by A.B.S., V.D., and J.K. within the
project AVOZ1-010-914 of the Academy of Sciences of Czech
Republic, and by L.H. as a part of the research program MSM 0021620837
of the Ministry of Education of Czech Republic. Financial support was
provided by the Grant Agency of the
Academy of Sciences of Czech Republic (Project A101100530), and
the Grant Agency of Czech Republic (Projects 202/04/1103 and 202/04/1005). We
gratefully acknowledge valuable discussions with W.E. Pickett, J.
Kune\v{s}, O. Eriksson, and G. Lander.

\end{document}